\documentclass[aps,amsmath,twocolumn,10pt]{revtex4-1}
\usepackage[english]{babel}
\usepackage{graphicx} 

\renewcommand{\vec}[1]{\mathbf{#1}}
\newcommand\rv{\vec{r}}

\newcommand\Hv{\vec{H}}
\newcommand\w{\omega}
\newcommand\eps{\varepsilon}

\newcommand\Vt{V'}
\newcommand\Jt{J'}

\begin{document}

\title{Topological Edge States in Bichromatic Photonic Crystals}

\author{F. Alpeggiani}
\author{L. Kuipers}

\affiliation{Kavli Institute of Nanoscience Delft, Department of Quantum Nanoscience, Delft University of Technology, Lorentzweg 1, 2628 CJ Delft, The Netherlands}

\begin{abstract}
Bichromatic photonic crystal structures are based on the coexistence of two different periodicities in the dielectric constant profile. They are realized starting from a photonic crystal waveguide and modifying the lattice constant only in the waveguide region. In this work, we numerically investigate the spectral and topological properties of bichromatic structures. Our calculations demonstrate that they provide a photonic analog of the integer quantum Hall state, a well known example of a topological insulator. The nontrivial topology of the bandstructure is illustrated by the formation of strongly localized, topologically protected boundary modes when finite-sized bichromatic structures are embedded in a larger photonic crystal.
\end{abstract}

\maketitle

\section{Introduction}

The integer quantum Hall effect (IQHE) has been the source of intense fascination since its discovery in 1980 \cite{Klitzing1980}. The integer values of the Hall conductance plateaus unveil a deep connection between the microscopic and the macroscopic world \cite{Kane2010}. When particles on a plane are subject to an external magnetic field, the competition between two different length scales, the lattice spacing of the crystalline potential and the magnetic length (the radius of the lowest-energy classical cyclotron orbit), gives rise to an intriguing energy spectrum with self-similarity characteristics, widely known as the Hofstadter butterfly \cite{Hofstadter1976}. Systems following an analogous physics to the Hofstadter model have been realized with arrays of microwave scatterers \cite{Stockmann1998}, ultracold atoms \cite{Jaksch2003,Ketterle2013,Bloch2013}, and in Van der Waals heterostructures \cite{Ponomarenko2013,Dean2013,Hunt2013}. A crucial property of the IQHE is that the Hall conductance is a topological invariant of the system, since its value is fully determined by the topology of the spectral structure in momentum space \cite{TKNN1982}. For this reason, the IQHE is an emblematic example of a topological insulator phase, i.e., a system that, albeit being insulating in the bulk, possesses topologically protected conducting states on the surface \cite{Kane2010}.

Recently, topological concepts have been transferred from the realm of electrons to that of light and various mechanisms to generate nontrivial topological states in photonic systems have been put forward \cite{review2014,review2018}. Early on, it was predicted that photonic crystals (PhCs) could give rise to topologically protected states when time-reversal symmetry is broken \cite{Haldane2008}. Unfortunately, the lack of suitable nonreciprocal materials in the optical range has restricted the experimental verification of the proposal to the microwave regime \cite{Wang2008}. In order to overcome such limitations, efforts have been focused on systems with external temporal modulation \cite{Fang2012,Minkov2016}, optomechanical devices \cite{Schmidt2015}, and photonic analogs of the quantum spin-Hall effect \cite{Hafezi2011,Carusotto2011,Hafezi2013,Rechtsman2013}. The latter approach has been also extended to photonic crystal systems \cite{Wu2015,Waks2016,Waks2018}.

A different route to evidence nontrivial topological phases in photonics is to investigate systems that are governed by the Harper-Aubry-Andr\'{e} (HAA) Hamiltonian \cite{Harper1955,AA1980}, which is the one-dimensional momentum-space projection of the IQHE and which inherits its nontrivial topological properties \cite{Chen2012,Zilberberg2012,DasSarma2013,Brouwer2013,Chong2015}. This approach has been experimentally carried out with cold atoms \cite{Inguscio2008} and coupled waveguide arrays, where topological pumping has been demonstrated \cite{Silberberg2009,Zilberberg2012,Zilberberg2013}. The same concept can be generalized to higher-dimensions and used to realize an analog of the four-dimensional IQHE \cite{Kraus2013,Lohse2018,Zilberberg2018}.

In this work, we investigate a novel class of photonic-crystal-based nanostructures which display nontrivial topological properties, called \emph{bichromatic structures}. These structures are based on the idea of an effective ``bichromatic potential'' for light, i.e., a spatial distribution of the dielectric index characterized by the superposition of two different periodicities. They were originally introduced as a strategy for realizing high-quality-factor optical cavities \citep{Alpeggiani2015}, with quality factors of the order of $10^6$ having been experimentally measured \cite{Simbula2017,DeRossi2017,DeRossiCLEO}. These cavities also provide a potentially viable route for reducing the power threshold of four-wave mixing and other parametric interactions \cite{DeRossi2017}. Here, we theoretically demonstrate that bichromatic structures embody the same physics of the HAA model, and, therefore, provide a photonic analog of the IQHE. The nontrivial topology of the optical spectrum is illustrated by the formation of strongly localized edge states at the boundary of the system. Thus, bichromatic structures represent a promising platform for investigating nontrivial states of light using state-of-the-art photonic crystal devices.

\section{Harper-Aubry-Andr\'{e} model}\label{sec:HAA}

\begin{figure}
\centering
\includegraphics[width=\linewidth]{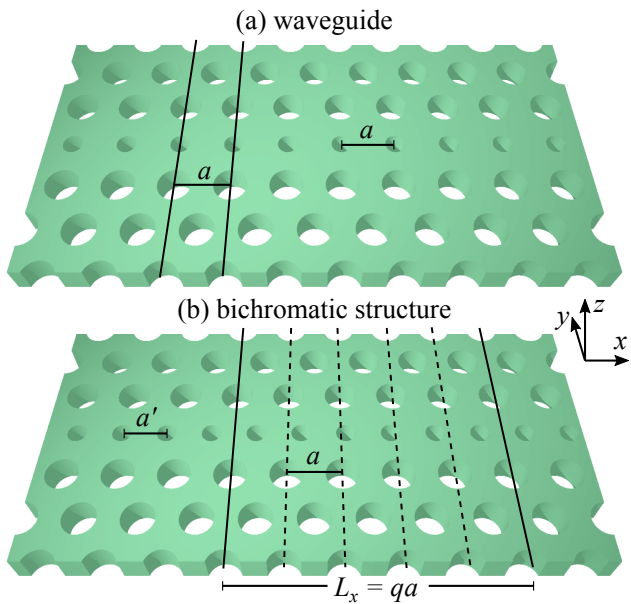}
\caption{Schematics of a standard PhC-slab waveguide (a) and a bichromatic PhC structure (b). In the bichromatic structure, the separation between the reduced-radius holes is modified to the value $a' = (q/p) a$ ($a$ is the lattice constant). In the example, $q = 5$ and $p = 6$. The solid lines mark the boundaries of the unit cells along the $x$ axis. The dashed lines in (b) delimit the one-dimensional lattice sites inside the supercell.}
\label{fig:schematic}
\end{figure}

The IQHE represents one of the simplest examples of a topological insulator. Although the bulk material is insulating, conduction is allowed along the boundaries of a finite-size sample. The existence of the conducting edge states is protected against deformations of the Hamiltonian (due to, for instance, disorder) by the conservation of a topological invariant, on condition that these deformations do not close the bulk bandgap \cite{Kane2010}.

In this section, we summarize some important results on the IQHE \cite{Kane2010,TKNN1982,Zilberberg2012}. In a typical description, nonrelativistic spinless particles (with charge $-e$) moving on a two-dimensional square lattice in the presence of a homogeneous magnetic field $H$ are considered. The magnetic field is directed perpendicularly to the lattice plane. We indicate with $\psi_{n,m}$ the Wannier wavefunction centered at the lattice site $(x,y) = (na,ma)$, with $a$ being the lattice constant. The Schr\"odinger equation for the particles reads \cite{Hofstadter1976}
\begin{multline}\label{eq:IQHE}
t(\psi_{n+1,m} + \psi_{n-1,m}) \\
+ t'(e^{i2\pi\beta n}\psi_{n,m+1}+
e^{-i2\pi\beta n}\psi_{n,m-1}) = \mathcal{E}\psi_{n,m},
\end{multline}
where $\mathcal{E}$ is the energy eigenvalue, $t$ and $t'$ the hopping terms along the $x$ and $y$ axes, respectively, and $\beta = ea^2 H / (2\pi\hbar c)$. When the particles hop along the $y$ axis, they acquire the additional phase term $\pm 2\pi\beta n$, which originates from the Peierls substitution with the vector potential $\vec{A} = (0,Hx,0)^T$. It is essential to note that the phase term is proportional to $n$, and, therefore, breaks the original periodicity along the $x$ axis.

The properties of the solutions of Eq.~\eqref{eq:IQHE} have been the subject of extensive investigation \cite{Hofstadter1976,TKNN1982,Kohmoto1985,Chen2012,Zilberberg2012,DasSarma2013,Brouwer2013,Chong2015}. Equation \eqref{eq:IQHE} gives rise to a set of bands ($\alpha = 1,2,\dots$), each associated with a topological invariant, the Chern number
\begin{equation}
C_{\alpha} = \frac{1}{2\pi i} \int_{\text{MBZ}} dk_1 dk_2 \left[\frac{\partial \mathcal{A}_2(\vec{k})}{\partial k_1} - \frac{\partial \mathcal{A}_1(\vec{k})}{\partial k_2}\right],
\end{equation}
where $\mathcal{A}_j(\vec{k}) = \langle \psi_{\alpha}(\vec{k})|\partial/\partial k_j |\psi_{\alpha}(\vec{k})\rangle$ is the Berry connection and the integral is performed over the magnetic Brillouin zone \cite{TKNN1982,Suzuki2005}. The existence of the Chern number imparts a topological structure onto the spectrum, which unveils itself when two different topological phases enter in contact. In this case, topologically protected edge states are formed, which are spatially localized at the boundary between the two different phases \cite{Chen2012,Zilberberg2012,Zilberberg2013}. An example of this effect can be observed at the edges of the sample, when the system is in a nontrivial topological phase ($C_{\alpha} \ne 0$).

It is straightforward to prove that, by Fourier transforming the wavefunctions of Eq.~\eqref{eq:IQHE} along the $y$ axis, the following equivalent equation is obtained:
\begin{equation}\label{eq:HAA}
t(\psi_{n+1} + \psi_{n-1}) + 2t'\cos(2\pi \beta n + \phi)\psi_n = \mathcal{E}\psi_{n},
\end{equation}
with $\phi = k_y a$. Equation~\eqref{eq:HAA} describes a one-dimensional chain of interacting particles in the presence of a periodic potential and it is usually called the Harper-Aubry-Andr\'{e} model \cite{Harper1955,AA1980}. As it has been recently demonstrated in pioneering works \cite{Chen2012,Zilberberg2012}, every one-dimensional physical system that is described by the model in Eq.~\eqref{eq:HAA} inherits the nontrivial topological properties of the integer quantum Hall effect. Notably, the existence of nontrivial Chern numbers and topologically protected edge states is not a property of the Hamiltonian for a single realization of the one-dimensional system, but of the entire set of one-dimensional Hamiltonians spanned by the phase shift $\phi$. The phase $\phi$ of the periodic modulation of the potential accounts for the momentum along the geometrical dimension that was lost when moving from a two-dimensional to a one-dimensional system.

\section{Bichromatic photonic crystals}
\label{sec:bichromatic}

The \emph{bichromatic} PhC structures that we study in this work are based on the superposition of two different periodicities in the dielectric index profile. This concept can be realized in practice by starting from a standard PhC waveguide and modifying the lattice constant only in the waveguide region, from the original value $a$ to a modified value $a'$ (Fig.~\ref{fig:schematic}).

We are interested in the the electromagnetic eigenmodes, i.e., the solutions of the second-order eigenvalue equation for the magnetic field \cite{Joannopoulos2008}
\begin{equation}\label{eq:eigen}
\nabla \times \left[\frac{1}{\eps(\rv)}\nabla\times \Hv(\rv)\right] = \frac{\w^2}{c^2}\Hv(\rv).
\end{equation}
We solve Eq.~\eqref{eq:eigen} using a guided-mode expansion method, where the magnetic field is expanded onto the basis of guided modes of a homogeneous dielectric slab with an effective dielectric index \cite{Andreani2006} (see the Supplementary Material). In this way, the complexity of the calculation is effectively reduced from a three-dimensional to a two-dimensional problem. Although this method is intrinsically approximate, since it neglects the interaction with the continuum of radiating modes, its efficacy for the study of photonic crystal structures is supported by a large body of work \cite{Andreani2006,GME2004,GME2010,GME2014,GME2018}.

\begin{figure}
\centering
\includegraphics[width=\linewidth]{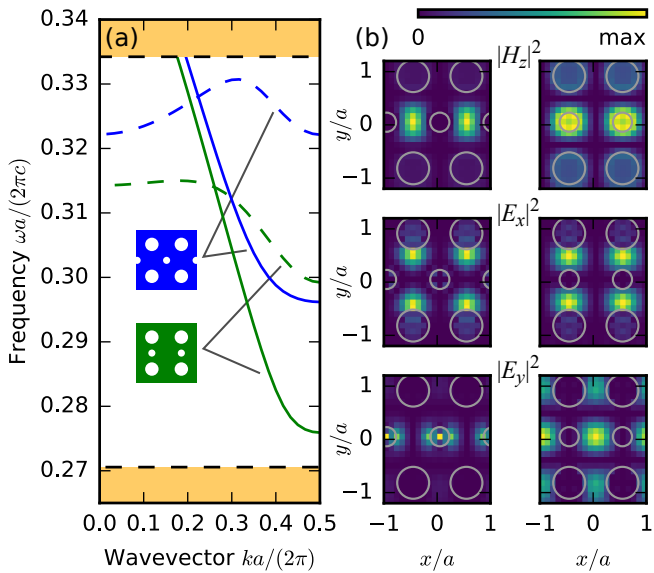}
\caption{(a) The frequency dispersion of two different configurations of a PhC waveguide (see insets), as a function of the wavenumber along the propagation direction. The solid and dashed lines represent modes where $H_z$ is even or odd by reflection along the $y$ axis, respectively. The shaded regions illustrate the approximate boundaries of the original PhC bandgap. (b) The intensity profiles of the $H_z$, $E_x$, and $E_y$ components of the field for the even modes in the two different configurations. The fields are computed at the edge of the first Brillouin zone ($k = \pi/a$).}
\label{fig:waveguide}
\end{figure}

The starting point for realizing bichromatic structures is a PhC waveguide. We consider a PhC slab with a triangular lattice of holes (with lattice constant $a$) patterned in a suspended high-refrective-index membrane. In this work, we will consider only TE-like modes, for which a bandgap opens in the spectrum \cite{Joannopoulos2008}. For the sake of illustration, we assume a silicon membrane ($\eps = 12$) of thickness $t = 0.5a$ and with hole radius $r = 0.3a$. The waveguide (linear defect) is realized by reducing the radius of the holes in a single row to the value $r_w = 0.18a$ [Fig.~\ref{fig:schematic}(a)]. In this way, additional wave\-guide modes appear inside the original PhC bandgap. In Fig.~\ref{fig:waveguide}(a), we plot the frequency dispersion of the waveguide modes for two different configurations: one in which the reduced-radius holes follow the pattern of the triangular lattice (blue curves, see inset), and another in which they are globally shifted by half lattice constant (green curves, see inset). For both configurations, we observe two different TE-like modes. These modes can be classified according to the even (solid curves) or odd (dashed curves) mirror-symmetry of the $H_z$ field with respect to the $xz$ plane. Note that the families of modes cannot couple with each other as long as the dielectric profile remains symmetric along the $y$ axis. In the following, we will focus only on the \emph{even} modes, which span the largest extension of the original PhC bandgap. The intensity of the electric and magnetic fields for the even modes of both configurations are illustrated in Fig.~\ref{fig:waveguide}(b) for the wavenumber $k = \pi/a$. It is interesting to note that the field profile is very similar in the two different configurations. However, in the shifted-row configuration there is a larger fraction of the electric field intensity localized in the silicon region. This fact corresponds to a more energetically favourable configuration and it results in globally lower modal frequencies, as shown in Fig.~\ref{fig:waveguide}(a). Furthermore, the dispersion of the even mode for the shifted-row configuration has a higher curvature at the band edge than the standard configuration. This behavior can be interpreted in terms of a smaller ``effective mass'' for the wave\-guide modes in the shifted-row configuration.

The variation in the dispersion of the waveguide modes with the global shift of the linear defect plays a central role for understanding the behavior of the bichromatic PhC structures. Bichromatic structures are realized by starting from the PhC waveguide that we just described and modifying the distance between the reduced-radius holes along the $x$ axis, from the original value $a$ to a different value $a'$ [Fig.~\ref{fig:schematic}(b)]. The crucial parameter is the ratio $\beta = a'/a$. Here, we consider structures with $\beta$ a rational number, i.e., $\beta = q/p$, with $p$ and $q$ coprime integer numbers. These structures are effectively periodic along the $x$ axis with a supercell of size $L_x = qa$. Inside the supercell, there are $p$ reduced-radius holes. The supercell boundaries for an illustrative system with $\beta = 5/6$ are indicated in Fig.~\ref{fig:schematic}(b). The behavior for irrational values of $\beta$ can be inferred from the limit of a series of commensurate structures with increasingly large supercells.

\begin{figure*}
\centering
\includegraphics{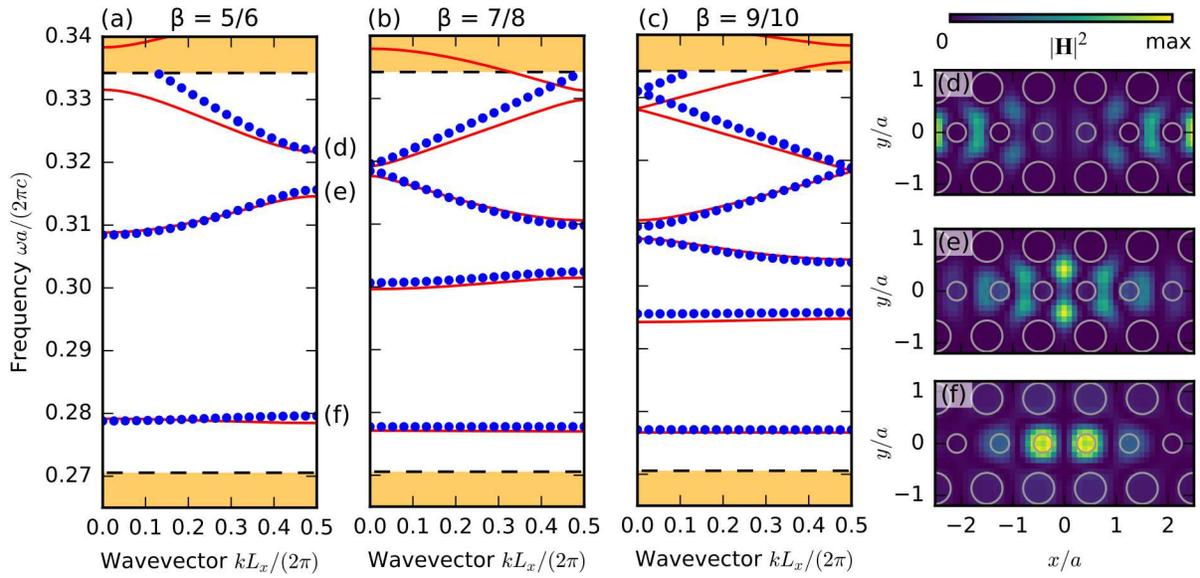}
\caption{The frequency dispersion of the $y$-even TE-like modes for three different examples of bichromatic structures: (a) $\beta = 5/6 = 0.8\overline{3}$, (b) $\beta = 7/8 = 0.875$, (c) $\beta = 9/10 = 0.9$. The dispersion is plotted inside one-half of the first Brillouin zone associated the one-dimensional supercell. The dots represent the frequency eigenvalues computed with the guided-mode expansion method, whereas the solid line are the eigenvalues of the model in Eq.~\eqref{eq:HAA2}, with the parameters in Eqs.~\eqref{eq:param1} and \eqref{eq:param2}. (d-f) The intensity of the magnetic field inside the supercell, for the three modes tagged in panel (a) at $k = \pi/L_x$.}
\label{fig:bandstructure}
\end{figure*}

In Fig.~\ref{fig:bandstructure}(a-c), we plot the modal frequencies (blue dots) of the $y$-even TE-like electromagnetic modes for several bichromatic structures with various values of $\beta$, as a function of the one-dimensional wavevector along the $x$-axis, $k$. Each plot displays one half of the one-dimensional Brillouin zone, whose absolute size is defined according to the effective supercell of the system [$-\pi/(qa) < k \le \pi/(qa)$] and, therefore, varies for each example. In all cases, however, we observe a striking difference with respect to the periodic waveguide of Fig.~\ref{fig:waveguide}. Instead of a continuous dispersion, we recognize a set of electromagnetic minibands, all lying inside the bandgap of the original photonic crystal slab. The number of the minibands depends on the size of the effective supercell, and, ultimately, on the integer $q$.

The origin of this behavior lies in the HAA model. As illustrated in Fig.~\ref{fig:schematic}(b), every lattice site inside the supercell realizes a different local configuration for the one-dimensional waveguide, which interpolates between the two limiting cases considered in Fig.~\ref{fig:waveguide}. Therefore, the waveguide mode will experience a different effective potential and mass at every lattice site. It is natural, then, to expand the magnetic field over a set of  Wannier functions centered at the lattice-site positions $x = na$ ($n = 1,2,\dots,q$), i.e., $\Hv(\rv) = \sum_n c_n \Hv_n(\rv)$ \cite{Alpeggiani2015}. In the nearest-neighbor approximation, the eigenvalue problem in Eq.~\eqref{eq:eigen} reduces to that for a one-dimensional chain of $q$ interacting particles
\begin{equation}
\frac{\w^2}{c^2}c_n = V_n c_n + J_n c_{n+1} + J_{n-1} c_{n-1},
\end{equation}
with site-dependent potential and hopping terms ($V_n$ and $J_n$, respectively). We assume periodic boundary conditions at the two ends of the finite-size chain with the Bloch wavenumber $k$, $-\pi/(qa) < k \le \pi/(qa)$. The additional periodicity $a' = (q/p)a$ implies a modulation of the potential and hopping terms which, neglecting higher-order harmonics, is proportional to $\cos(2\pi x/a' + \phi) = \cos(2\pi np/q + \phi)$. Thus, the eigenvalue problem takes the form of the generalized one-dimensional HAA model \cite{DasSarma2013}
\begin{equation}\label{eq:HAA2}
\begin{aligned}
\frac{\w^2}{c^2}c_n &= \left[V + \Vt\cos(2\pi pn/q + \phi_V)\right]c_n\\
&+ \left[J + \Jt \cos(2\pi pn/q + \phi_J)\right]c_{n+1}\\
&+ \left[J + \Jt \cos(2\pi p(n-1)/q + \phi_J)\right]c_{n-1}.
\end{aligned}
\end{equation}
This model represents a straightforward generalization of the one in Eq. \eqref{eq:HAA}, and it is characterized by equivalent topological properties \cite{Zilberberg2012, DasSarma2013}.

We observe that it is possible to accurately fit the frequency dispersion of bichromatic structures within a simple version of the generalized HAA model where only the off-diagonal modulation is present ($\Vt = 0$). The results of this fit are shown by the red lines in Figs.~\ref{fig:bandstructure}(a-c). The fitting parameters will be discussed below. It is noteworthy that this simple model correctly reproduces the number of minibands, their relative energy separation, and their main curvature. The decrease of accuracy for higher frequencies is likely due to the hybridization with the continuum of modes above the photonic-crystal bandgap. The choice of the fitting parameters is likely not unique, and a comparable or even better agreement with the full-wave simulations might be obtained by including higher-harmonic modulation terms and higher-order hopping constants in the HAA model of Eq.~\eqref{eq:HAA2}. However, the increased complexity goes beyond the scope of the present analysis, which is to provide a minimal model for understanding the topological behavior of bichromatic structures. In particular, our results confirm that bichromatic structures provide a practical realization of the HAA model. Therefore, in the light of our previous discussion, they can be thought to represent a photonic analog of the integer quantum Hall effect.

The electromagnetic modes of bichromatic structures can be interpreted as ``Landau levels'' for an effective two-dimensional system. The ratio $\beta = a'/a$ then plays the role of the effective magnetic flux. This interpretation is supported by the field profiles of the electromagnetic modes, a selection of which is illustrated in Figs.~\ref{fig:bandstructure}(d-f). The lowest-frequency mode [Fig.~\ref{fig:bandstructure}(f)] is strongly localized around the most energetically favourable PhC waveguide configuration. Higher-frequency modes have a larger effective volume and stretch towards the edges of the supercell. Note that, in the limit $\beta \lesssim 1$, i.e., $q \lesssim p$ and $q, p \gg 1$, the size of the supercell might become comparable with the actual size of the sample. In this case, owing to their mostly flat dispersion [see Figs.~\ref{fig:bandstructure}(a-c)], the lowest-order electromagnetic Bloch modes effectively become localized nanocavity modes, with a field distribution similar to the one in Fig.~\ref{fig:bandstructure}(f). The initial theoretical investigation of bichromatic structures \cite{Alpeggiani2015} and recent experimental works \cite{Simbula2017,DeRossi2017,DeRossiCLEO} have been mostly focused only on this subset of electromagnetic modes, treated as localized cavity states.

\section{The spectrum of bichromatic structures}

\begin{figure}
\centering
\includegraphics{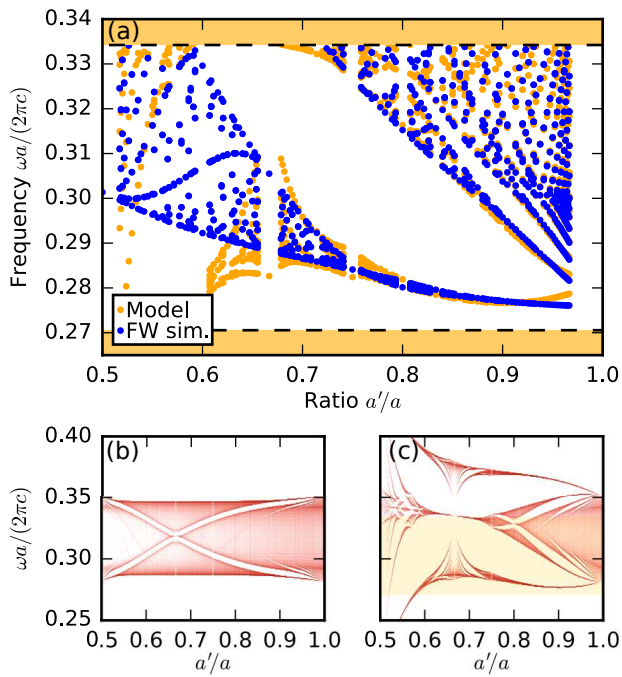}
\caption{(a) The spectrum of bichromatic structures as a function of $\beta = a'/a$, computed at the band-edge point $k = \pi/L_x$. The blue dots are the results of a full-wave (FW) simulation with the guided-mode expansion method, whereas the orange dots are computed from the model in Eqs.~\eqref{eq:HAA2}, \eqref{eq:param1}, and \eqref{eq:param2}. The latter are also depicted in panel (c), over a broader frequency range (the shaded region corresponds to the original PhC bandgap). For comparison, panel (b) illustrates the spectrum of the model in Eq.~\eqref{eq:HAA2} assuming that the parameters are constant over the range of variation of $\beta$.}
\label{fig:spectrum}
\end{figure}

One of the most intriguing properties of the spectrum of bichormatic structures for rational $\beta$ is that the number of minibands is determined by the integer denominator $q$. In the HAA model of Eq.~\eqref{eq:HAA2}, the actual number of solutions is exactly given by $q$. In the case of the bichromatic structures, however, the actual number of accessible solutions is reduced to the ones lying inside the bandgap of the embedding photonic crystal. In both cases, since a small change in $\beta$ can produce a huge variation in $q$, the number of minibands wildly fluctuates as a function of $\beta$.

Motivated by these considerations, we investigate the spectrum of bichromatic structures as a function of $\beta$. The results of full-wave electromagnetic simulations are summarized in Fig.~\ref{fig:spectrum}. The blue dots in Fig.~\ref{fig:spectrum}(a) represent the eigenfrequencies of the Bloch modes at $k = \pi/L_x$ for the rational values of $\beta = a'/a = q/p$ ($q \le 30$). The intricate structure of the spectrum bears a striking resemblance with the well known Hofstadter's butterfly \cite{Hofstadter1976}. Such similarity is not surprising, as the Hofstadter's butterfly illustrates the spectrum of the original HAA model in Eq.~\eqref{eq:HAA}. However, in order to accurately describe the spectrum of bichromatic structures, we have to take into account some additional physical effects. In fact, by changing the value of the ratio $\beta$, the air-dielectric fraction of the linear defect and the magnitude of the nearest-neighbor hopping are also altered. These effects can be accounted for in the model of Eq.~\eqref{eq:HAA2} by assuming $\beta$-dependent parameters $V(\beta)$, $J(\beta)$, and $\Jt(\beta)$. We find that the guided-mode simulation results of Fig.~\ref{fig:spectrum}(a) can be fitted with the model in Eq.~\eqref{eq:HAA2} and the linear-dependent parameters
\begin{gather}\label{eq:param1}
a^2 V = 5.30 - 1.30\beta,\: \Vt = 0 \\\label{eq:param2}
a^2 J = 1.03 - 0.66\beta,\: a^2 \Jt = 1.79 - 1.73\beta.
\end{gather}
These parameters are specific to the present choice of the PhC geometry and are governed by the effective potential and effective mass of the photons in the waveguide. However, the justification of the HAA model lies in the coexistence of different periodicities. Therefore, we expect that the same model will describe bichromatic structures with different geometrical configurations, albeit with different sets of parameters. Further details about the fitting procedure are presented in the Supplementary material.

The effect of the additional $\beta$-dependence of the parameters can be understood by comparing Figs.~\ref{fig:spectrum}(b) and \ref{fig:spectrum}(c). In Fig.~\ref{fig:spectrum}(b), we show an example of the spectrum of Eq.~\eqref{eq:HAA2} for constant parameters [corresponding to the values of Eqs.~\eqref{eq:param1} and \eqref{eq:param2} for $\beta = 1$], resulting in a highly symmetric structure. The additional $\beta$-dependence of the parameters induces a deformation of the spectral structure, which is illustrated in Fig.~\ref{fig:spectrum}(c). The same spectral points are depicted by the orange dots in Fig.~\ref{fig:spectrum}(a), in order to allow for an easier comparison with the full-wave simulation data. We find good agreement with the HAA model in the approximate range $0.75 \lesssim \beta \lesssim 0.96$. The agreement deteriorates for lower values of $\beta$. This effect is likely due to the increased deviation of the local field profile from that of the original waveguide in Fig.~\ref{fig:waveguide}, undermining the validity of the tight-binding model in Eq.~\eqref{eq:HAA2}.

These results confirm the existence of a range for the parameter $\beta = a'/a$ where bichromatic PhC structures truly embody the physics of the HAA model, providing a photonic analog of a topological insulator. The additional $\beta$-dependence of the parameters of the model can be interpreted as a dependence of the potential and hopping terms on the effective magnetic field, inducing a deformation of the spectrum with respect to the original Hofstadter's model \cite{Hofstadter1976}. Similar spectral deformations have also been found for realistic condensed-matter systems, due to, for instance,  diamagnetic effects \cite{Hernandez2013}.

\section{Topological edge states}

\begin{figure*}
\centering
\includegraphics[scale=0.95]{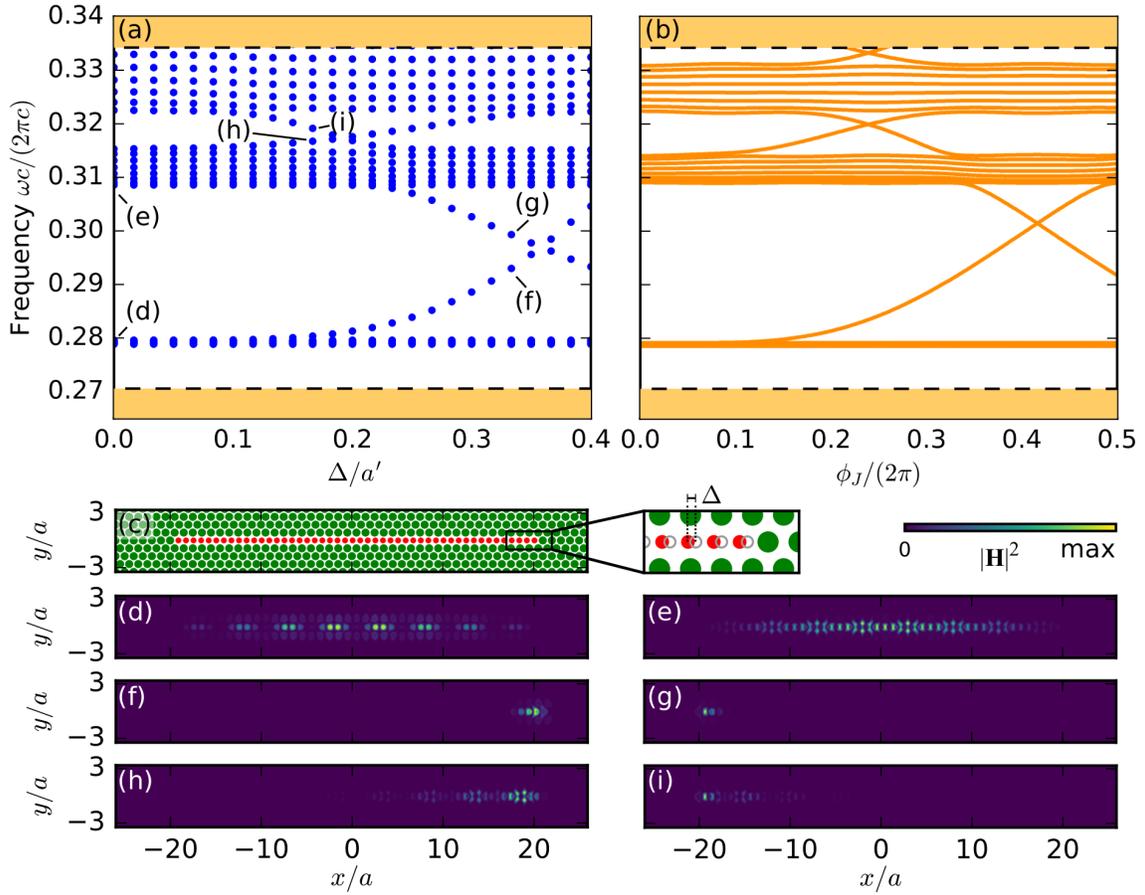}
\caption{(a) Full-wave calculation of the frequency eigenvalues of a finite-sized bichromatic structure as a function of the spatial displacement $\Delta$. We assume $\beta = 5/6$ and a finite-sized extent of $N_r = 8$ repetitions of the bichromatic supercell. (b) Eigenvalues of a finite-sized chain of particles following the HAA model of Eqs.~\eqref{eq:HAA2}, \eqref{eq:param1}, and \eqref{eq:param2} for the same parameters, as a function of the phase shift $\phi_J$. (c) Schematics of the dielectric profile of the structure being simulated. The holes with radius $r = 0.3a$ are represented in green, whereas those with $r_w = 0.18a$ are shown in red. The close-up illustrates the effect of the global displacement $\Delta$. (d-i) The intensity of the magnetic field for various modes tagged in panel (a).}
\label{fig:edges}
\end{figure*}

As we summarized in Sec.~\ref{sec:HAA}, the HAA model is characterized by topologically nontrivial energy bands, each of which can be associated with a topologically invariant Chern number \cite{Kane2010,TKNN1982,Kohmoto1985}. For instance, in the case $\beta = 5/6$, we calculate from the model in Eq.~\eqref{eq:HAA2} that the Chern numbers associated to the bands are $C_1 = C_2 = C_4 = C_5 = 1$, and $C_3 = -4$ (with the index running from low to high energies) \cite{Suzuki2005}. Similarly, the Chern numbers for the other two examples of Fig.~\ref{fig:bandstructure} are $C_4 = -6$ and $C_5 = -8$ for $\beta=7/8$ and $\beta=9/10$, respectively, with $C_{\alpha} = 1$ for the remaining bands. As originally demonstrated in Refs.~\cite{Chen2012,Zilberberg2012}, a compelling manifestation of the topological structure of the HAA model is the formation of edge states between two spatially distinct topological phases. For instance, these edge states appear at the boundaries of the HAA system when it is enclosed in a topologically trivial medium.

We consider a large, yet finite-size, sample of a bichromatic structure, which encompasses a number of periodic repetitions of the bichromatic supercell. This finite-sized extent of the bichromatic system is  embedded inside a larger PhC slab, which plays the role of a topologically trivial region. For illustration, in Fig.~\ref{fig:edges} we summarize the results for the case $\beta = 5/6$. We assume $N_{r} = 8$ periodic repetitions, implying a total number $N_{h} = 48$ of reduced-radius holes, which are illustrated by red circles in the outline of the dielectric profile of Fig.~\ref{fig:edges}(c). The remaining standard-radius holes are represented by the green circles. As discussed in Sec.~\ref{sec:HAA}, in order to demonstrate the presence of edge states, we need to restore the additional geometrical dimension that is lost when we move from the two-dimensional IQHE [Eq.~\eqref{eq:IQHE}] to the one-dimensional HAA model of Eqs.~\eqref{eq:HAA} and \eqref{eq:HAA2}. The lost geometrical dimension is mapped onto the global phase shift of the periodic potential modulation [$\phi = k_y a$ in Eq.~\eqref{eq:HAA}]. In our photonic realization of the HAA model, this parameter corresponds to the global spatial displacement $\Delta$ of the line of reduced-radius holes with respect to the surrounding PhC. The geometrical meaning of the global displacement $\Delta$ is highlighted in the close-up of Fig.~\ref{fig:edges}(c).

We calculate the eigenvalues of the finite-size system in Fig.~\ref{fig:edges}(c) as a function of the displacement $\Delta$. The eigenvalue frequencies are displayed in Figs.~\ref{fig:edges}(a). Most of the eigenfrequencies lie inside the minibands investigated in Sec.~\ref{sec:bichromatic} [compare Fig.~\ref{fig:bandstructure}(a)], as expected for a finite-size section of a periodic system. The corresponding eigenmodes are delocalized all over the bichromatic region, as shown, for instance, by the intensity profiles in Figs.~\ref{fig:edges}(d,e), which illustrate examples of Bloch modes from the first and the second band, respectively.

However, an eye-catching feature is the presence of additional modes which cross the gaps among the minibands. The nature of these modes is immediately clear by looking at the intensity profile of the the magnetic field, which is displayed in Figs.~\ref{fig:edges}(f--i) for some selected values of $\Delta$. The modes are strongly localized at the edges of the finite-sized extension of the bichromatic structure. This behavior indicates that they are the photonic analog of the edge states that originate in the HAA model due to bulk-edge correspondence. In Fig.~\ref{fig:edges}(b), we plot the spectrum of a finite-dimensional chain of particles following the HAA Hamiltonian in Eq.~\eqref{eq:HAA2}. The agreement with the full-wave simulation results for the bichromatic strcuture [Fig.~\ref{fig:edges}(a)] is compelling, further supporting the physical analogy between the two systems. The relation with the edge states of the IQHE can be grasped by looking at the effective ``group velocity'' for the variable $\Delta$, i.e., $\partial\w/\partial\Delta$. In the light of Sec.~\ref{sec:HAA}, this quantity is the analog of the group velocity along the $y$-axis for the particles, i.e., $\partial\mathcal{E}/\partial k_y$. Therefore, the sign of the group velocity represents the direction of motion of a flux of particles moving along a finite-sized two-dimensional stripe of material. As shown in Fig.~\ref{fig:edges}(a), the effective group velocity is positive for edge states localized at the right of the sample, and vice versa. Thus, the states at the two opposite edges can be associated with effective currents of particles flowing in opposite directions. Note that the exact same result applies to the case $\Delta < 0$ (which is the mirror symmetric situation along $x$), since both the effective group velocity and the position of the edge states are reversed.

The two lowest-frequency minibands in Fig.~\ref{fig:edges}(b) have the same Chern number ($C_1 = C_2 = 1$), which is different from the Chern number of the higher-frequency band ($C_3 = -4$). Therefore, we expect that the edge states will connect the two lowest bands with the upper one. This behavior is indeed suggested by the intensity distribution of the edge modes. The edge states in Figs.~\ref{fig:edges}(g,i) have a similar profile, despite lying in different minigaps. In both cases, the field profile along the $y$ axis resembles the one of the lowest-frequency miniband [compare Fig.~\ref{fig:edges}(d)] and considerably differs from that of the edge state in Fig.~\ref{fig:edges}(h), which is more similar to the second miniband [Fig.~\ref{fig:edges}(e)]. Thus, the properties of the edge modes reflect the global topology of the bandstructure.

The results presented in this section refer to a system with $\beta = 5/6$. We have also computed the spectrum of finite-sized bichromatic structures with different values of $\beta$, observing in all cases the formation of the strongly localized modes at the edges of the structures. Similarly to the present example, these modes lie inside the minigaps and appear for specific values of the displacement $\Delta$. Additional spectra for different values of $\beta$ are presented in the Supplementary Material.

\section{Conclusion}

In this work, we investigate the properties of a novel class of nanophotonic systems, bichromatic PhC structures. We theoretically demonstrate that bichromatic structures provide a realization of the HAA model. Therefore, they also exhibit a photonic analog of the integer quantum Hall state, a well known example of a topological insulator. The nontriviality of the bandstructure topology is evidenced by the formation of spatially-localized edge modes when a finite-sized bichromatic structure is embedded in a larger PhC. These electromagnetic modes are analogous to the  topologically protected edge states of the IQHE.

It is important to note that the geometrical configuration of bichromatic structures is not specifically tailored to provide the sinusoidal modulation of the effective potential and mass required by the HAA Hamiltonian. Rather, the modulation naturally emerges from the superposition of two different periodicities. The situation is different from other optical systems, for instance coupled waveguide arrays, where the HAA Hamiltonian is realized by explicitly tuning the width or the position of each waveguide \cite{Kraus2013}. Moreover, bichromatic structures are characterized by a single essential degree of freedom, the ratio $\beta$ between the two periodicities. Starting from a different PhC waveguide configuration, as done for instance in Ref.~\cite{DeRossi2017}, will result in a similar modulation of the effective potential for light, provided that two competing periodicities are present in the system.

For these reasons, bichromatic structures represent a promising platform for visualizing topological effects in PhC systems. For instance, bichromatic structures could serve as a basis for realizing topological pumping of light \cite{Kraus2013} across a photonic device. Moreover, the concept can be generalized to investigate nontrivial states of light in higher-dimensions \cite{Kraus2013,Lohse2018,Zilberberg2018} or in the presence of time-modulated optical properties, for instance in optomechanical or nonlinear systems.

\section*{Funding Information}
H2020 Marie Sk\l{}odowska-Curie Actions (individual fellowship BISTRO-LIGHT, No. 748950); FP7 Ideas: European Research Council (ERC Advanced Grant No. 340438-CONSTANS).


\bibliography{bib}

\end{document}